\documentclass[conference]{IEEEtran}
\usepackage{graphicx}
\usepackage{caption}
\graphicspath{ {images/} }
\usepackage[font=small]{subfig}
\usepackage[font=small,labelfont=bf]{caption}
\usepackage[english]{babel}
\usepackage[table]{xcolor}
\usepackage{algorithmic}
\usepackage{algorithm}
\usepackage{amsmath}
\usepackage{amsmath}
\usepackage[a4paper,left=0.63in,right=0.63in,top=0.75in,bottom=2.95cm]{geometry}
\usepackage{url}
\usepackage{balance}
\begin{document}

\title{Incorporation of Confidence Interval into Rate Selection Based on the Extreme Value Theory for Ultra-Reliable Communications\\
}


\author{\IEEEauthorblockN{Niloofar Mehrnia\IEEEauthorrefmark{1} 
and Sinem Coleri\IEEEauthorrefmark{2}
}
\IEEEauthorblockA{\IEEEauthorrefmark{1}Koc University Ford Otosan Automotive Technologies Laboratory (KUFOTAL), Sariyer, Istanbul, Turkey, 34450}
Email: nmehrnia17@ku.edu.tr\\
\IEEEauthorblockA{\IEEEauthorrefmark{2}Department of Electrical and Electronics Engineering, Koc University, Istanbul, Turkey, 34450}
Email: scoleri@ku.edu.tr
}

\maketitle

\begin{abstract}
Proper determination of the transmission rate in ultra-reliable low latency communication (URLLC) needs to incorporate a confidence interval (CI) for the estimated parameters due to the large amount of data required for their accurate estimation.
In this paper, we propose a framework based on the extreme value theory (EVT) for determining the transmission rate along with its corresponding CI for an ultra-reliable communication system. This framework consists of characterizing the statistics of extreme events by fitting the generalized Pareto distribution (GPD) to the channel tail, deriving the GPD parameters and their associated CIs, and obtaining the transmission rate within a confidence interval.   
Based on the data collected within the engine compartment of Fiat Linea, we demonstrate the accuracy of the estimated rate obtained through the EVT-based framework considering the confidence interval for the GPD parameters. Additionally, we show that proper estimation of the transmission rate based on the proposed framework requires a lower number of samples compared to the traditional extrapolation-based approaches.

\end{abstract}


\section{Introduction}
\label{sec:intro}
Ultra-reliable low latency communication (URLLC) is a substantial new feature brought by the fifth-generation ($5$G), with a potential to support a vast set of machine type and mission-critical applications such as reliable remote control of robots, tele-operation driving (ToD) applications, or coordination among vehicles \cite{interface_01}-\nocite{MehrniaTWC}\cite{urllc_01}. A system operating at URLLC is expected to address the strict reliability requirement of packet error rate (PER) at $10^{-9}$-$10^{-5}$ level, while the tolerable latency is on the order of milliseconds or less \cite{interface_01}-\nocite{5G_01}\nocite{interface_02}\nocite{urllc_02}\nocite{MehrniaTWC}\cite{MehrniaTVT}. Fulfillment of the reliability requirements at URLLC necessitates a statistical model of the channel tail distribution \cite{MehrniaTWC}, \cite{MehrniaTVT}, as well as the transmission strategies guaranteeing ultra-reliability concerns \cite{urllc_05}. An accurate estimation of the wireless channel parameters in the ultra-reliable regime requires a massive amount of data. However, analyzing the confidence intervals in the estimated parameters as a function of the amount of historical data allows a significant reduction in the minimum sample numbers required for the estimation of wireless channel tail distribution at URLLC.

Current studies on the statistical modeling of the wireless channel for ultra-reliable communications either extrapolate a wide range of commonly used practical channel models toward the ultra-reliability region \cite{urllc_02}, \cite{urllc_05}, or utilize extreme value theory (EVT) to characterize the tail distribution of the channel data and derive the statistics of extreme events efficiently with a minimum amount of data \cite{MehrniaTWC}, \cite{MehrniaTVT}. However, none of these studies assess the reliability of the system through the estimation of the transmission rate. 

Measuring the system reliability and the transmission rate for ultra-reliable communications has been addressed in the context of new performance measures of the reliability based on the transmission rate \cite{urllc_05}, and proposing a rate selection framework in the absence of knowledge about the channel state information (CSI) \cite{reliability_03}. 
In \cite{urllc_05}, two performance measures have been proposed with the goal of selecting a transmission rate guaranteeing ultra-reliability: average reliability (AR) for dynamic environments where the channel changes frequently; and probably correct reliability (PCR) for the static environments where the channel statistics can be considered constant for an extensive time interval. Moreover, \cite{reliability_03} evaluates if the training times are affected by the cumulative distribution function (CDF) of the fading channel when a perfect CSI is not available, through estimating an optimal transmission rate for a channel and then, measuring the reliability of the system accordingly.  
Nevertheless, such approximations characterize the tail statistic based on the extrapolation of the average-statistics channel models, and therefore, might be deficient to fulfill the reliability targets of ultra-reliability \cite{MehrniaTWC}.
Only recently, we have proposed a novel EVT-based framework for the estimation and validation of the optimal transmission rate for ultra-reliable communications \cite{MehrniaTVTRate}. However, the possible changes in the transmission rate due to the estimation error have been neglected in the computation.

The reliability of the system is strongly affected by the accuracy of the estimated channel parameters due to their effect on the transmission rate. Therefore, confidence interval (CI) is derived as a metric to guarantee the reliability constraint despite the limited amount of training data.
In \cite{confidenceinterval_03}, the authors propose a non-parametric statistical learning algorithm that utilizes kernel density estimation (KDE) for estimating the probability density function (PDF) of the channel distribution and then, upon selecting a suitable transmission channel rate, evaluates the outage probability within a confidence level to maintain the URLLC reliability. However, the proposed empirical non-parametric approach is based on the extrapolation of the average statistics fading channels.
The authors in \cite{reliability_01} proposed a data-driven learning framework to characterize the statistics of a non-blocking connectivity duration within a confidence interval, in the absence of statistical knowledge about the dynamic wireless channel, with the goal of addressing the ultra-reliability constraint in URLLC. However, the machine learning data-driven approaches require an extensive large amount of training samples, much more than ($10 \times \epsilon^{-1}$) to meet the targeted error probability $\epsilon$.
 
The goal of this study is to propose a novel framework based on EVT for the estimation of the optimal transmission rate along with the corresponding confidence interval with a sufficiently low number of samples for ultra-reliable communications. According to the proposed framework, upon incorporating the estimates of the generalized Pareto distribution (GPD) parameters fitted to the channel tail into the rate selection, the upper and lower bounds of the GPD parameters are employed to acquire the confidence interval for the transmission rate. The original contributions of the paper are listed as follows: 

\begin{itemize}
    \item We propose a methodology based on EVT for ultra-reliable communications, where the channel tail distribution is modeled by using GPD, the upper and lower bounds for the estimated Pareto parameters are derived and then incorporated into the estimation of the confidence interval for the transmission rate.
    \item We determine the confidence intervals of the estimated GPD parameters by incorporating $t$-distribution and multiple iterations for the parameter estimation using the maximum likelihood estimator (MLE). 
    \item We formulate the confidence interval of the transmission rate in conjunction with the confidence interval of the estimated GPD parameters.
    \item Based on the data collected within the engine compartment of Fiat Linea, we apply our methodology at different numbers of training samples to ensure that the targeted error probability of URLLC is met even when limited data is available.
\end{itemize}

The rest of the paper is organized as follows. Section \ref{sec:system_model} describes the system model and assumptions considered throughout the paper. Section~\ref{sec:methodology} presents the EVT-based framework for determining the rate selection and its affiliated confidence interval in connection with the accuracy of the GPD parameters fitted to the tail distribution of the channel.
Section \ref{sec:numerical_results} provides the channel measurement setup, and the performance evaluation in determining the confidence interval for the estimated transmission rate. Finally, concluding remarks and future works are given in Section \ref{sec:conclusions}.

\section{System Model}
\label{sec:system_model}
We consider a single transmitter (Tx)-receiver (Rx) pair communicating over a stationary channel, i.e., the parameters of the channel distribution are fixed over a vast period of time. If the channel is non-stationarity according to the Augmented Dickey-Fuller (ADF) test results, the external factors causing time variation in the parameters of the GPD are determined such that the sequence is divided into $M$ groups, each of which can be considered stationary, as explained in details in \cite{MehrniaTVT}.
The transmit power is assumed to be fixed and known in advance. Therefore, estimating the received signal power is equivalent to estimating the squared amplitude of the channel state information \cite{urllc_05}, \cite{MehrniaTWC}.

We assume that Tx sends a packet to an Rx at rate $R$ over an unknown channel. Applying an EVT-based channel modeling methodology, the statistics of the channel tail distribution are estimated by fitting the generalized Pareto distribution to the received power values exceeding a given threshold. Afterward, the confidence intervals of the estimated parameters corresponding to different probabilities of wrong decisions are derived for different numbers of samples. Then, the statistics of the transmission rate are estimated according to the EVT by incorporating the confidence intervals of the estimated Pareto parameters for different sample numbers to meet the targeted error probability at an ultra-reliable communication. 


\section{EVT-based Rate Selection Framework}
\label{sec:methodology}
We propose an EVT-based framework to model the channel tail distribution of the received powers by using GPD, determine the transmission rate guaranteeing ultra-reliability, and estimate the confidence interval for the estimated Pareto parameters and the transmission rate. The framework consists of the following steps: The received power sequence is converted into independent and identically distributed (i.i.d.) samples since EVT requires i.i.d. sequence as an input. Then, GPD is fitted to the tail of i.i.d. received powers to model the channel tail distribution. Upon estimating the Pareto parameters by using MLE, the optimum transmission rate for an ultra-reliable communication is determined. Finally, the confidence intervals of the estimated GPD parameters and the transmission rate are computed at different sample numbers. By incorporating the confidence interval into our proposed framework, we expect to decrease the number of samples required to determine the optimum rate that meets ultra-reliable constraints in URLLC. 
 
\subsection{Channel Estimation}
\label{sec:channelestimation}
EVT provides a robust framework for analyzing the statistics of extreme events happening rarely through modeling the probabilistic distribution of the values exceeding a given threshold by using the GPD \cite{MehrniaTWC}, \cite{evt_04}. To determine the statistics of the channel tail distribution, prior to the transmission and at the training phase, Rx collects stationary channel measurements and then, convert them into the sequence of $n$ i.i.d. samples, denoted by $X^n = \{x_1, ...,x_n\}$, with the cumulative distribution function (CDF) $F$, where $x_{i}$ is the $i^{th}$ sample, $i \in \{1,...,n\}$. Upon converting the measured samples into an i.i.d. sequence by removing their dependency using the declustering approach \cite{MehrniaTWC}, the Generalized Pareto Distribution (GPD) is fitted to the received power values exceeding a given threshold to model the channel tail distribution as
\begin{equation}
\label{eqn:gpddist}
    F_{u}(y) = 1-\big[1+\frac{\xi y}{\tilde{\sigma}_{u}}\big]^{-1/\xi},
\end{equation}
where $y$ is a non-negative value denoting the exceedance below threshold $u$, i.e., $y=u-X | X=\{x_i<u\}$; $\xi$ and $\tilde{\sigma}_{u}=\sigma+\xi(u-\mu)$ are shape and scale parameters of the GPD, respectively; and $\mu$ and $\sigma$ are the location and scale parameters of the generalized extreme value (GEV) distribution fitted to the CDF of $M_n = max \{ x_1,...,x_n\}$, respectively \cite[Theorem~1]{MehrniaTWC}, \cite{evt_11}. 

The optimum threshold $u$ is determined by utilizing two complementary methods, mean residual life (MRL) and parameter stability methods \cite{MehrniaTWC}, \cite{evt_04}, while the estimation of Pareto distribution parameters is performed by using the Maximum Likelihood Estimation (MLE). The validity of the estimated Pareto model is then assessed by using probability plots including probability/probability (PP) plot and quantile/quantile (QQ) plot.

\subsection{Rate selection}
\label{sec:rateselection}
The optimum transmission rate, assuming the unit bandwidth, is determined such that:
\begin{equation}
    \label{eqn:maxrate}
    R_{GPD}(X^n) = \log_{2} \big(1+u + \frac{\hat{\sigma}}{\hat{\xi}}\big[1 - \varepsilon_{n}^{-{\hat{\xi}}} \big]\big),
\end{equation}
where $R_{GPD}(X^n)$ denotes the transmission rate determined based on $n$ training received power samples; $u$ is the optimum threshold for the channel tail estimation; $\hat{\sigma}$ and $\hat{\xi}$ are the MLE of the scale and shape parameters of the GPD($\sigma$,$\xi$) fitted to the tail distribution of the training samples $X^{n} = \{x_1,x_2,...,x_n\}$, respectively; and $\varepsilon_n$ is the outage probability calculated with respect to the targeted PER $\epsilon$, as follows:
\begin{equation}
    \label{eqn:en}
    \varepsilon_n = \big[ 1-\frac{\hat{\xi}}{\hat{\sigma}} \big(\frac{\sigma}{\xi}\big(1-\epsilon^{-\xi}\big)\big)\big]^{-\frac{1}{\hat{\xi}}},
\end{equation} 
where $\xi$ and $\sigma$ are the shape and scale parameters of the GPD, given that $F_u(.)$ is perfectly known and GPD is fitted to the sample sequence including the data in both training and test phases \cite{MehrniaTVTRate}.
Then, the lower and upper bounds of the estimated Pareto parameters, as well as the estimated transmission rate are calculated for different values of probability of wrong decision $\alpha$ and at different sample numbers. The small $\alpha$ refers to a critical value based on which, we are $100(1-\alpha)\%$ confident that the estimated parameter is within the derived confidence interval.

\subsection{Confidence Interval}
\label{sec:cicalculation}
\subsubsection{CI for the estimated Pareto parameters}
\label{sec:ciparam}
The estimated scale and shape parameters of GPD are not exactly $\hat{\sigma}$, $\hat{\xi}$ and typically take values within the confidence interval. Suppose that we have estimated the value of the GPD parameter $\theta(i)$ for a given realization $i$, where $\theta(i)$ is either scale ($\sigma$) or shape ($\xi$) parameter. Our goal is to determine $(1-\alpha)$ confidence interval, $[l(i),u(i)]$, such that 
\begin{equation*}
    Pr\big(E(\theta_i) \in \big[l(i),u(i)\big]\big) \geq 1-\alpha,
\end{equation*}
for a small value of $\alpha \in (0,1)$ \cite{confidenceinterval_02}, where 
\begin{equation}
\label{eqn:upperlowerbound}
    \begin{array}{ll}
     l(i) = \bar{\theta}_i - t^{*} s_{i}, \\
     u(i) = \bar{\theta}_i + t^{*} s_{i}
\end{array}
\end{equation}
are the lower and upper bounds of the CI, respectively. In (\ref{eqn:upperlowerbound}), $t^{*}$ is the critical value obtained from a $t$-distribution with $M-1$ degrees of freedom, $M$ denotes the iteration number to estimate the parameter; and $s_{i}$ is the standard deviation of the $\theta_i$, calculated by $s_{i} =\sqrt{({1}/{M}) \sum_{j=1}^{M}(\theta_j-\bar{\theta}_i)^{2}}$ while $\bar{\theta}_i$ is the expected value of parameter $\theta$ derived by taking an average of the estimated values over $M$ iterations. Assuming a small quantity, such as $0.01$ or $0.05$ for the critical value $\alpha$, we are $100(1-\alpha)\%$ confident that the estimated parameter $\theta$ is within the range $[l(i),u(i)]$. 

\subsubsection{CI for the transmission rate}
\label{sec:cirate}
According to (\ref{eqn:maxrate}), the transmission rate of a channel operating at URLLC is a function of the GPD parameters. Since changes in the estimated Pareto parameters alter the transmission rate, the upper and lower bounds of the GPD parameters yield corresponding upper and lower bounds for the transmission rate. Referring to (\ref{eqn:maxrate}), the upper/lower bound of scale parameter $\hat{\sigma}$ along with the lower/upper bound of the shape parameter $\hat{\xi}$ provide the upper/lower bound of the transmission rate $R_{GPD}(X^n)$. Someone might say that $R_{GPD}(X^n)$ is also a function of $\varepsilon$. However, by assuming large number of training samples on the order of $1/\epsilon$, $\varepsilon_n \approx \epsilon$. It is due to the fact that the very small difference between the actual GPD parameters and the estimated ones are negligible and therefore, $(\frac{\hat{\xi}}{\hat{\sigma}})(\frac{\sigma}{\xi})$ and $(-\xi)(-\frac{1}{\hat{\xi}})$ in the power of (\ref{eqn:en}), are $\approx 1$.
As a result, the only parameters affecting the upper and lower bounds of the transmission rate in (\ref{eqn:maxrate}), are the lower and upper bounds of the estimated GPD parameters fitted to $n$ training samples, i.e., $\hat{\xi}$ and $\hat{\sigma}$. Accordingly, the lower/upper bound of the transmission rate is calculated as follows:
\begin{equation}
    \label{eqn:rxci}
    R_{GPD}(X^n)_{l/u} = \log_{2} \big(1+u + \frac{\hat{\sigma}_{l/u}}{\hat{\xi}_{u/l}}\big[1 - \varepsilon_{n}^{-{\hat{\xi}_{u/l}}} \big]\big),
\end{equation}
where subscription $l/u$ denotes the lower/upper bound of the confidence interval for the parameters already defined in (\ref{eqn:maxrate}).

\section{Numerical Results}
\label{sec:numerical_results}
The goal of this section is to evaluate the performance of the proposed framework in modeling the tail distribution of the channel at URLLC, estimating the transmission rate of the channel, and determining the confidence interval of the rate through the computation of the confidence interval for the parameters of GPD fitted to the tail distribution of the channel data, for different probabilities of wrong decision, $\alpha$ values, and different sample numbers.

\begin{figure}[h]
\centering
\captionsetup[subfigure]{labelformat=empty}
     \begin{center}
        \subfloat[(a)]{%
            \label{fig:engin}
            \includegraphics[width=0.45\columnwidth]{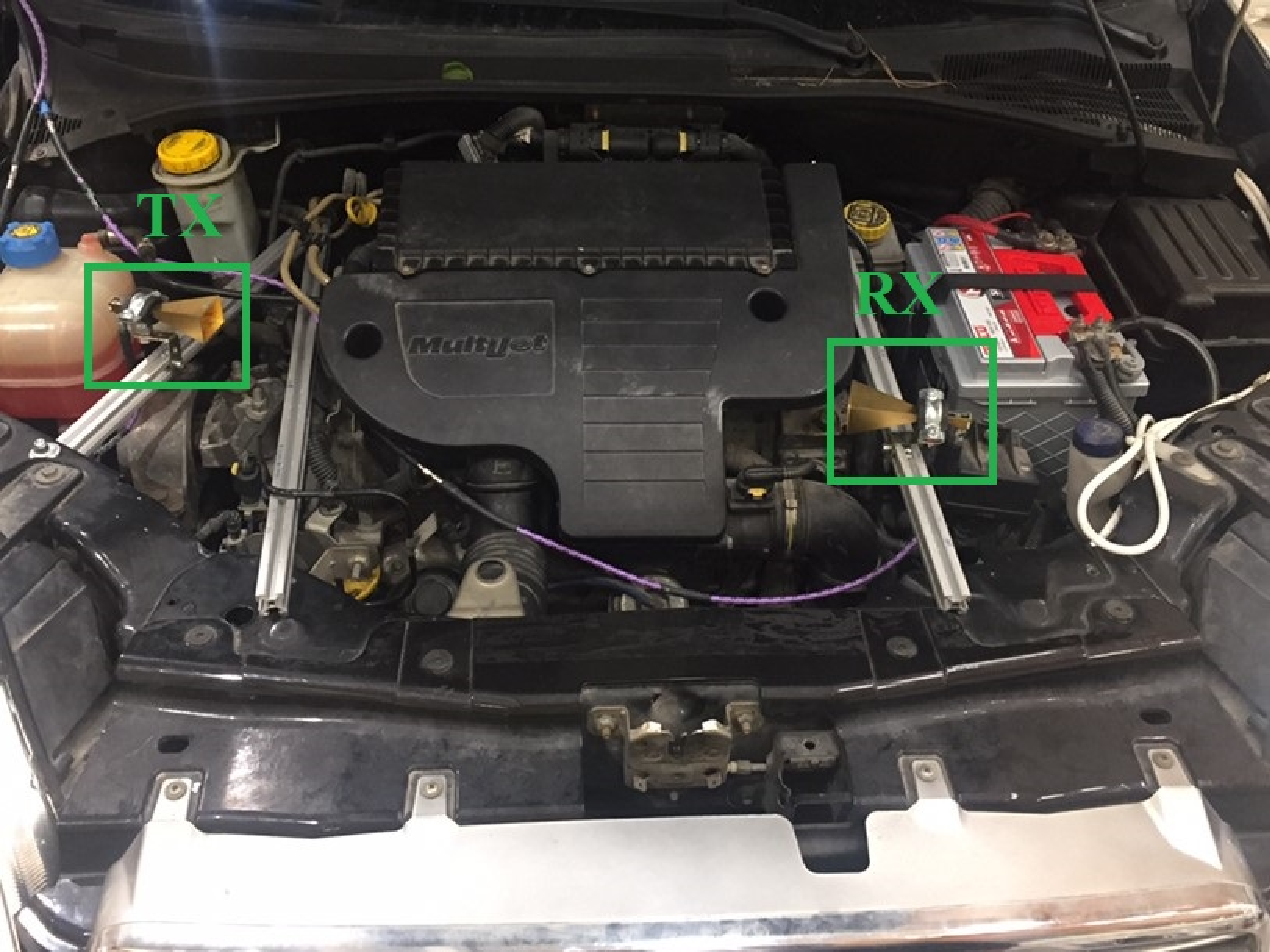}
        }%
        \subfloat[(b)]{%
            \label{fig:vna}
            \includegraphics[width=0.45\columnwidth]{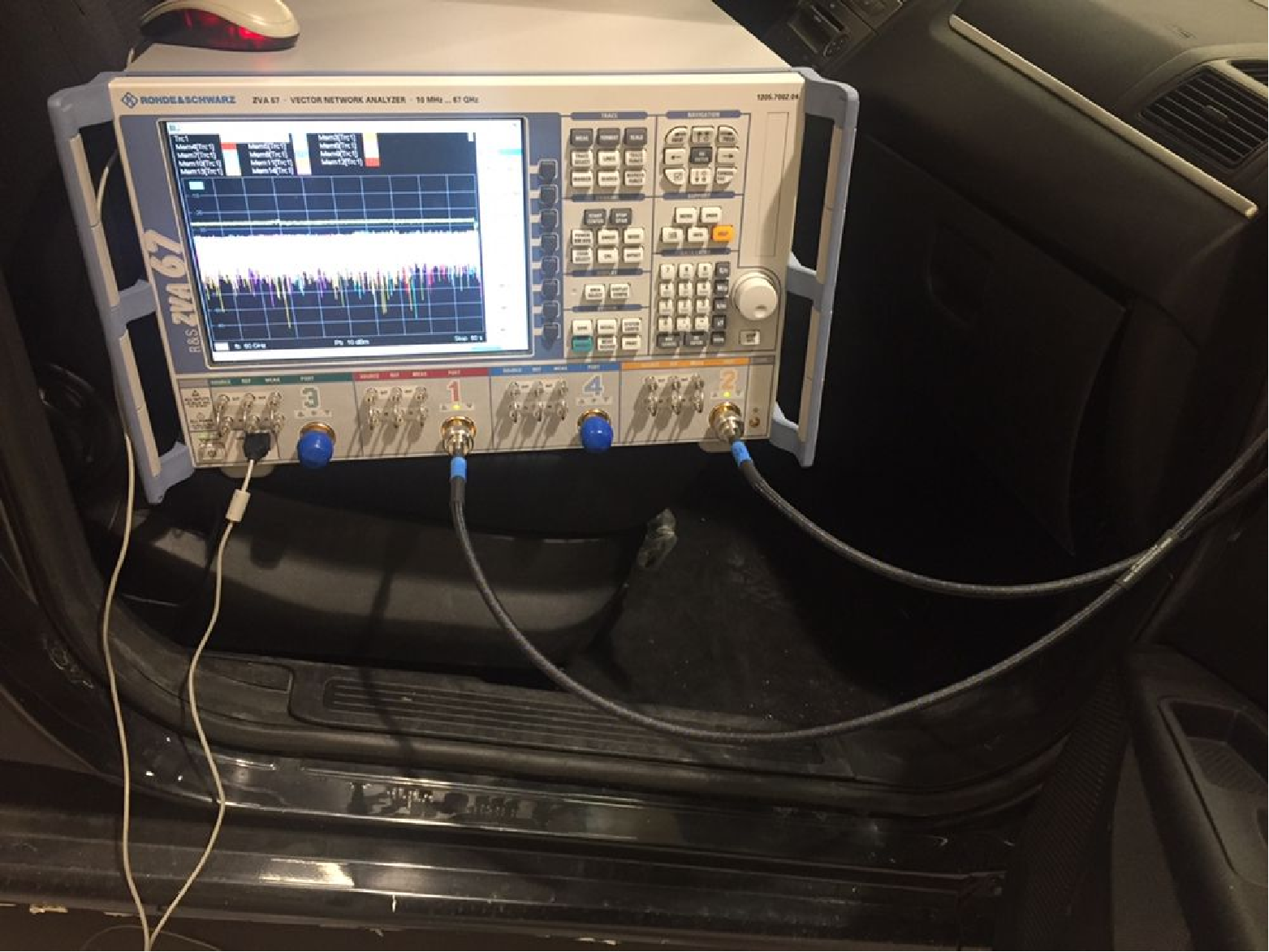}
        }\\ 
    \end{center}
    \vspace{-3mm}
    \caption{Measurement setup with the transmitter (TX) and receiver (RX) antennas located in the engine compartment of Fiat Linea: (a) Engine compartment, and (b) VNA setup.}
   \label{fig:probplotsArima}
\end{figure}

The channel data were collected within the engine compartment of Fiat Linea at $60$ GHz. The implemented antennas at the engine compartment are located such that the effect of the engine vibration is observed in the received power, as shown in Fig.~\ref{fig:engin}. A Vector Network Analyzer (VNA) (R$\And$S$\textsuperscript{\textregistered}$ ZVA$67$) is connected to the transmitter and receiver via the R$\And$S$\textsuperscript{\textregistered}$ ZV-Z$196$ port cables with maximum $4.8$ dB transmission loss, as shown in Fig.~\ref{fig:vna}. The horn transmitter and receiver antennas with a nominal $24$ dBi gain and $12^\circ$ vertical beam-width operate at $50$-$75$ GHz. About $10^{6}$ successive samples have been captured for $30$ minutes with a time resolution of $2$ ms. 

\begin{figure}[t!]
\centering
\captionsetup[subfigure]{labelformat=empty}
     \begin{center}
        \subfloat[(a)]{%
            \label{fig:ciscaleg1}
            \includegraphics[width=0.65\columnwidth]{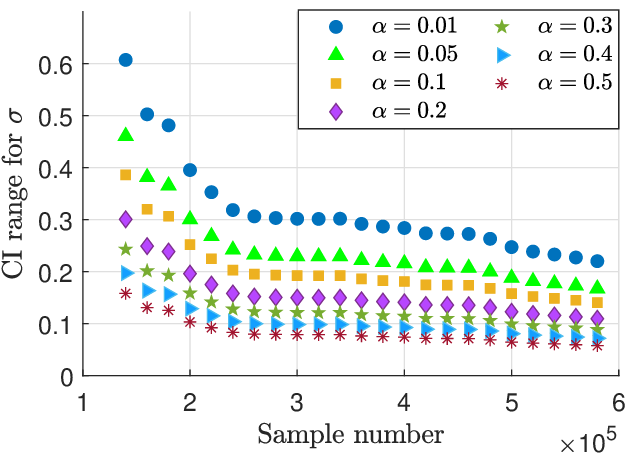}
        }\\
        \subfloat[(b)]{%
            \label{fig:cishapeg1}
            \includegraphics[width=0.65\columnwidth]{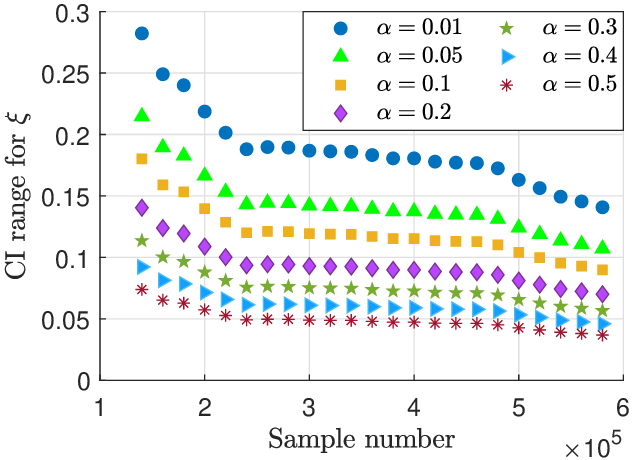}
        }\\
        \subfloat[(c)]{%
            \label{fig:ciscaleg2}
            \includegraphics[width=.65\columnwidth]{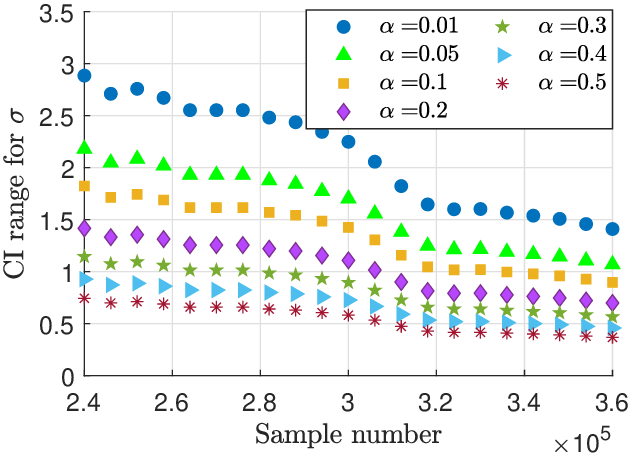}
        }\\
        \subfloat[(d)]{%
            \label{fig:cishapeg2}
            \includegraphics[width=0.65\columnwidth]{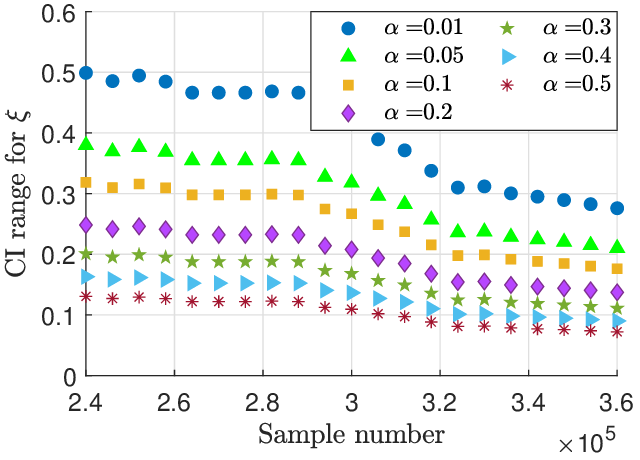}
        }
        \\
    \end{center}
    \caption{CI range for the estimated Pareto parameters considering different $\alpha$ values for groups $1$ and $2$ at different sample numbers: (a) CI range of the scale parameter for group $1$, (b) CI range of the shape parameter for group $1$, (c) CI range of the scale parameter for group $2$, and (d) CI range of the shape parameter for group $2$. The CI range refers to the difference between the upper and lower bounds of the CI.}
   \label{fig:ciparam}
\end{figure}
\begin{figure}[h!]
\centering
\captionsetup[subfigure]{labelformat=empty}
     \begin{center}
        \subfloat[(a)]{%
            \label{fig:paramscaleg1}
            \includegraphics[width=0.65\columnwidth]{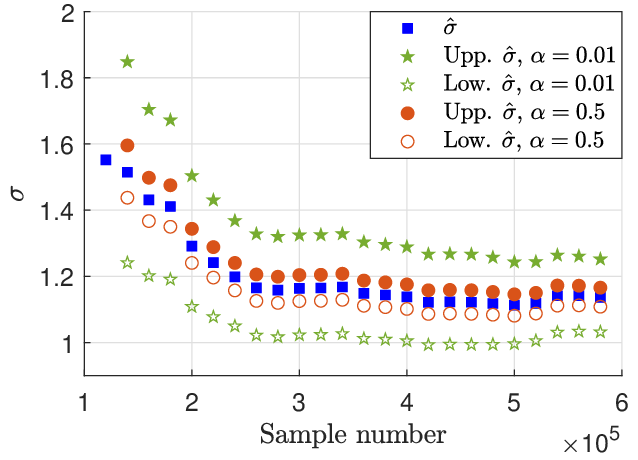}
        }\\
        \subfloat[(b)]{%
            \label{fig:paramshapeg1}
            \includegraphics[width=0.65\columnwidth]{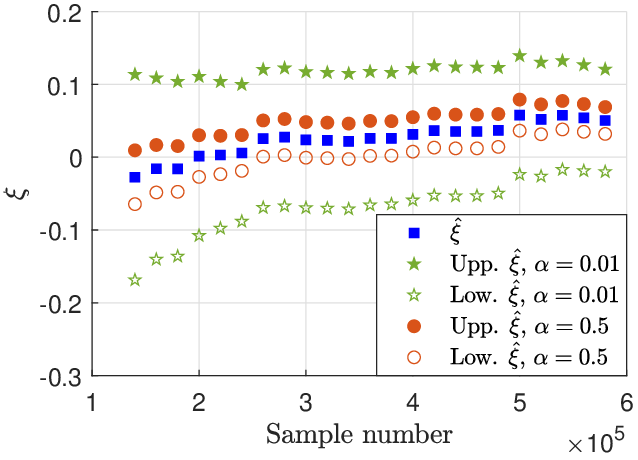}
        }\\
        \subfloat[(c)]{%
            \label{fig:paramscaleg2}
            \includegraphics[width=.65\columnwidth]{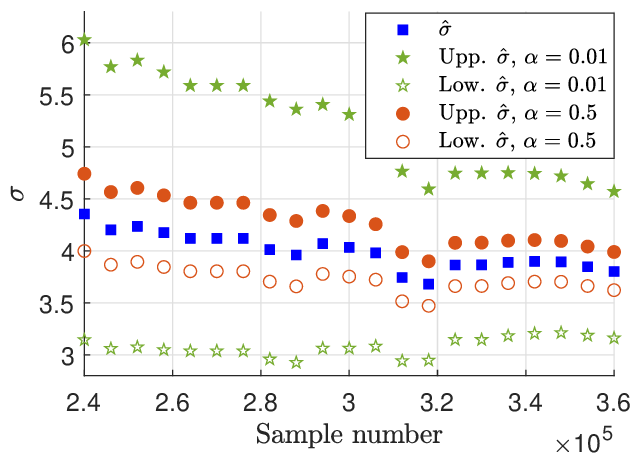}
        }\\
        \subfloat[(d)]{%
            \label{fig:paramshapeg2}
            \includegraphics[width=0.65\columnwidth]{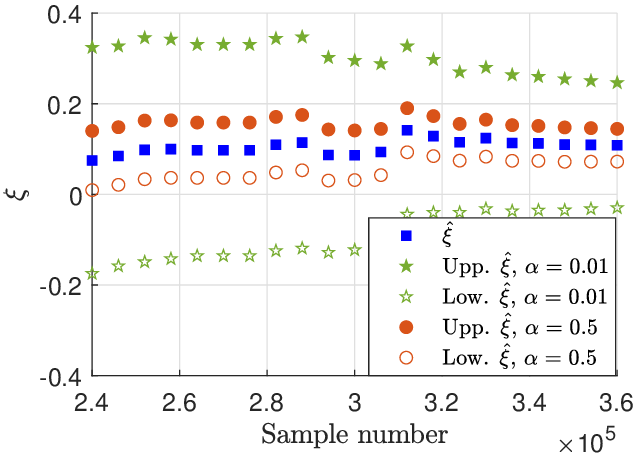}
        }
        \\
    \end{center}
    \caption{The estimated Pareto parameters along with their CI considering $\alpha=0.01,0.5$ for groups $1$ and $2$ at different sample numbers: (a) Scale parameter and the corresponding CI for group $1$, (b) Shape parameter and the corresponding CI for group $1$, (c) Scale parameter and the corresponding CI for group $2$, and (d) Shape parameter and the corresponding CI for group $2$. Blue plot is the estimated parameter; circle/pentagram corresponds to $\alpha=$ $0.5$/$0.01$; filled/empty plot corresponds to upper/lower bound of CI.}
   \label{fig:param}
\end{figure}

\textsc{MATLAB} is used for the implementation of the proposed framework. Due to the existence of a non-stationarity trend among the samples, the measured samples are categorized into two stationary groups according to the engine vibration. Group $1$ includes bunches of $10^3$ successive samples all above $-12$ dBm. If any sample within $10^3$ samples is less than $-12$ dBm, the set of successive samples is assigned to group $2$. It should be noted that the optimum threshold for group $1$ and group $2$ are estimated as $-5$ dBm and $-20$ dBm, respectively, and the targeted packet error rate, $\epsilon$ in (\ref{eqn:en}), is $10^{-5}$.

\subsection{CI for the GPD parameters}
\label{sec:numericalciparam}
Fig.~\ref{fig:ciparam} illustrates the confidence interval range for the estimated GPD parameters considering different $\alpha$ values at different sample numbers for groups $1$ and $2$. The range of confidence interval refers to the absolute value of the difference between the upper and lower bounds of the CI interval. As the $\alpha$ value increases from $0.01$ to $0.5$, i.e., allowing more error in the estimation of the GPD parameters, the CI range decreases significantly. 
Additionally, by increasing the number of training samples, the CI range decreases remarkably and, the estimation accuracy for the GPD parameters increases. Moreover, since group $2$ suffers from more severe extreme values, the minimum number of training samples required to estimate the GPD parameters fitted to the tail distribution is higher that that of group $1$.

Fig.~\ref{fig:param} shows the estimated GPD parameters along with the lower and upper bounds of CIs for $\alpha= \{0.01,0.5\}$ and different sample numbers in both groups $1$ and $2$. Please note that Fig.~\ref{fig:ciparam} depicts the CI range while Fig.~\ref{fig:param} illustrates the CI itself for the estimated GPD parameters. 
Above a certain number of samples, the estimated GPD parameters become almost constant and invariant with respect to the number of samples. Furthermore, the MLE with fewer samples yields a high level of uncertainty in the estimated scale ($\sigma$) and shape ($\xi$) parameters, but the uncertainty decreases as the sample number increases. This highlights the trade-off between the uncertainty of the GPD parameters and the cost of collecting data. Unlike the traditional non-parametric rate selection technique, which requires substantially more channel training than $1/\epsilon$ stated in \cite{urllc_05}, our EVT-based framework appropriately estimates the rate within a confidence interval with a sample size of roughly $1/\epsilon$.

\subsection{CI for the transmission rate}
\label{sec:numericalcirate}
\begin{figure}[t]
\centering
\captionsetup[subfigure]{labelformat=empty}
     \begin{center}
        \subfloat[(a)]{%
            \label{fig:cirateg1}
            \includegraphics[width=0.65\columnwidth]{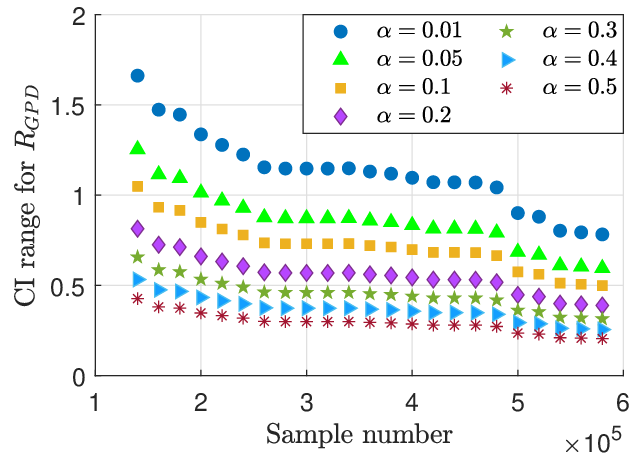}
        }\\
        \subfloat[(b)]{%
            \label{fig:cirateg2}
            \includegraphics[width=0.65\columnwidth]{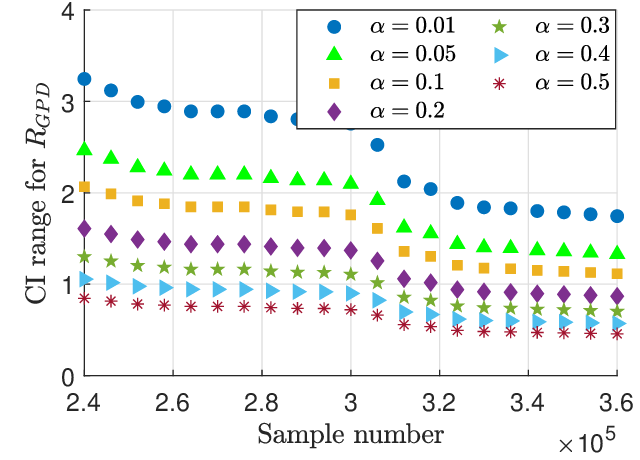}
        }\\ 
        
    \end{center}
    \caption{CI range for the estimated transmission rate considering different $\alpha$ values and different sample numbers for: (a) group $1$, and (b) group $2$.}
   \label{fig:cirate}
\end{figure}

Fig.~\ref{fig:cirate} illustrates the CI range for the transmission rate considering different $\alpha$ values at different sample numbers for groups $1$ and $2$. 
Similar to what we observed in Fig.~\ref{fig:ciparam}, increasing the $\alpha$ value reduces the CI range. In addition, increasing the sample size during the training phase dramatically decreases the CI range, while converging to a fixed value beyond a certain large enough sample number. Furthermore, the minimum sample numbers necessary to fit the GPD to the tail distribution of channel data are higher in group $2$ than in group $1$.

\begin{figure}[h!]
\centering
\captionsetup[subfigure]{labelformat=empty}
     \begin{center}
        \subfloat[(a)]{%
            \label{fig:rateg1}
            \includegraphics[width=0.65\columnwidth]{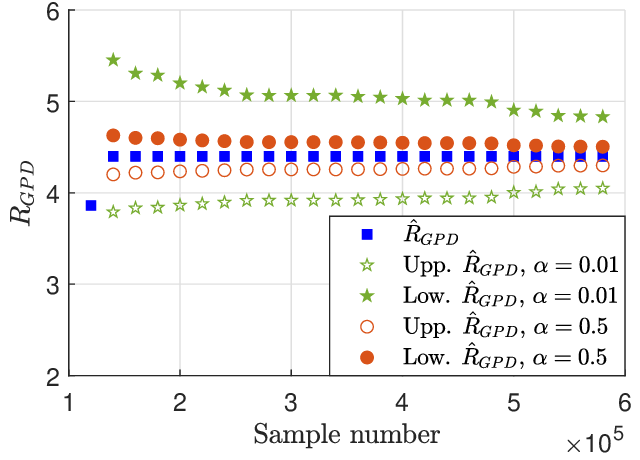}
        }\\
        \subfloat[(b)]{%
            \label{fig:rateg2}
            \includegraphics[width=0.65\columnwidth]{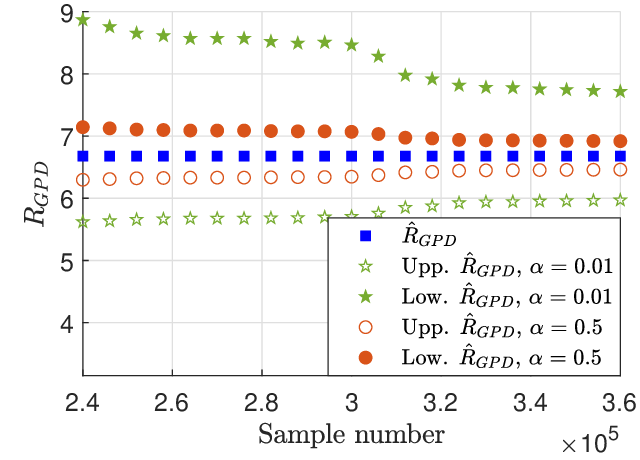}
        }\\ 
        
    \end{center}
    \caption{Estimated transmission rate with the corresponding CI at $\alpha = 0.5, 0.01$ and different sample numbers for: (a) group $1$, and (b) group $2$. Blue plot is the estimated parameter; circle/pentagram corresponds to $\alpha=$ $0.5/0.01$; filled/empty plot corresponds to upper/lower bound of CI.}
   \label{fig:rate}
\end{figure}

Fig.~\ref{fig:rate} depicts the estimated transmission rate with the corresponding CI for $\alpha = \{0.01, 0.5\}$ and different sample numbers for groups $1$ and $2$. Same as the difference between Figs.~\ref{fig:ciparam} and \ref{fig:param}, Fig.~\ref{fig:cirate} illustrates the CI range while Fig.~\ref{fig:rate} shows the CI itself for the estimated transmission rate.
The estimated transmission rate is constant with respect to the sample sizes. However, the confidence interval fluctuates with sample size, especially for lower $\alpha$ values. By incorporating a substantial large sample number in the training phase in both groups $1$ and $2$, the lower and upper boundaries of the confidence interval for the transmission rate become constant, and so the estimated transmission rate stays within an approximately fixed interval.

\section{Conclusions}
\label{sec:conclusions}
In this paper, we present a novel EVT-based methodology for calculating the transmission rate of a system operating in the ultra-reliable regime within its appropriate confidence interval. First, we estimate the generalized Pareto distribution (GPD) parameters fitted to the channel tail and then, we use these estimates along with their confidence intervals, to compute the transmission rate and its accompanying confidence interval. The incorporation of CI results into rate selection estimate reduces sample complexity by reducing the number of samples necessary for rate estimation in the training phase to attain a specific level of reliability. Furthermore, calculating the GPD parameters with few samples results in a wider CI in the rate estimation, indicating considerable uncertainty, but the uncertainty decreases as the sample number increases.
This emphasize the trade-off between the GPD parameters uncertainty, controlling the uncertainty in the rate selection, and the cost of data collection. In the future, we plan to extend the proposed framework for the multivariate EVT analysis by considering multiple Tx-Rx scenarios.

\balance 

\ifCLASSOPTIONcaptionsoff
  \newpage
\fi
\bibliographystyle{ieeetr}
\bibliography{CIParamEst.bib}

\end{document}